# JGR Space Physics

## Observations of Asymmetric Lobe Convection for Weak and Strong Tail Activity

A. Ohma[1], N. Østgaard[1], J. P. Reistad[1], P. Tenfjord[1], K. M. Laundal[1], T. Moretto Jørgensen[1], S. E. Haaland[1,2], P. Krcelic[2,3], and S. Milan[4]

[1]Birkeland Centre for Space Science, Department of Physics and Technology, University of Bergen, Bergen, Norway, [2]Max-Planck Institute for Solar Systems Research, Göttingen, Germany, [3]Department of Geophysics, University of Zagreb, Zagreb, Croatia, [4]Department of Physics and Astronomy, University of Leicester, Leicester, UK

**Abstract** In this study we use high-quality convection data from the Electron Drift Instrument on board Cluster to investigate how near-Earth tail activity affects the average convection pattern in the magnetotail lobes when the interplanetary magnetic field has a dominating east-west ($B_y$) component. Two different proxies have been used to represent different levels of reconnection in the near-Earth tail: The value of the AL index and the substorm phases identified by the Substorm Onsets and Phases from Indices of the Electrojet algorithm. We find that the convection changes from a dominantly $Y_{GSM}$ direction, but opposite in the two hemispheres, to a flow oriented more toward the plasma sheet, as the north-south component of the convection increases when reconnection enhances in the near Earth tail. This result is consistent with recent observations of the convection in the ionosphere, which suggest that the nightside convection pattern becomes more north-south symmetric when tail reconnection increases. This is also supported by simultaneous auroral observations from the two hemispheres, which shows that conjugate auroral features become more symmetric during substorm expansion phase. The reduced asymmetry implies that the asymmetric pressure balance in the lobes is altered during periods with strong reconnection in the near-Earth tail.

## 1. Introduction

The plasma circulation in the magnetosphere is primarily controlled by the solar wind and the interplanetary magnetic field (IMF). For a purely southward IMF, this circulation is described by the Dungey cycle (Dungey, 1961): The IMF can reconnect with the terrestrial field at the dayside magnetopause, and the field lines opened by this process are dragged across the polar caps by the solar wind and added to the magnetotail lobes. These field lines will eventually close again by reconnecting in the magnetotail and return to the dayside to repeat the cycle. This transport of magnetic flux sets up a two-cell convection pattern in the ionosphere in both hemispheres, with an antisunward convection of open field lines across the polar caps and a sunward return flow of closed field lines at lower latitudes, both at dusk and at dawn. Reconnection in the magnetotail can occur at different locations, either in the distant tail or in the near-Earth tail ($10-25\ R_E$). Reconnection in the distant tail closes open field lines directly. Near-Earth reconnection, however, is associated with substorms and commences at closed field lines but progresses to open field lines as the substorm evolves (e.g., Hones, 1979).

Research from the past decades have shown that the entire magnetospheric system becomes highly asymmetric when an east-west ($B_y$) component is present in the IMF. Numerous studies have shown that the two-cell convection pattern then consists of one round "orange" cell and one crescent "banana" cell, based on in situ observations from low-altitude spacecraft (e.g., Heppner & Maynard, 1987) and high-altitude spacecraft (e.g., Förster et al., 2007; Haaland et al., 2007), as well as ground based measurements (e.g., Pettigrew et al., 2010; Thomas & Shepherd, 2018). The pattern mirrors over the noon-midnight meridian between the Northern and Southern Hemispheres, which means that there is a north-south asymmetry. This asymmetry is also manifested in the plasma convection in the magnetotail lobes. Instead of mainly convecting toward the neutral sheet in the $Z_{GSM}$ direction, the convection velocity will have an oppositely directed dawn-dusk component in the Northern and Southern Hemispheres (Gosling et al., 1984, 1985; Haaland et al., 2008; Noda et al., 2003). The convection velocity and direction depend strongly on the IMF direction, and the magnitude of the convection is positively correlated with the magnitude of the IMF (Haaland et al., 2008, 2009).

The north-south asymmetry is also reflected in the orientation of the magnetic field in the closed magnetosphere. The field lines are displaced from their initial configuration, introducing a $B_y$ component in the closed magnetosphere with the same sign as the IMF $B_y$ component (Cowley & Hughes, 1983; Petrukovich, 2011; Wing et al., 1995) and displacing the footpoints of the field lines in opposite directions in the two hemispheres. This displacement is regularly observed in auroral images from the two hemispheres, which show a systematic longitudinal displacement of conjugate auroral features consistent with the prevailing IMF conditions (Frank & Sigwarth, 2003; Østgaard et al., 2004, Østgaard, Laundal, et al., 2011; Reistad et al., 2013, 2016). A consistent asymmetry is also observed in both the Birkeland currents and the ionospheric Hall and Pedersen currents (e.g., Anderson et al., 2008; Green et al., 2009; Laundal et al., 2016, 2018).

The asymmetry in the magnetosphere originates from the loading of flux from dayside and lobe reconnection, which, under the influence of an IMF $B_y$ component, is added asymmetrically to the two lobes (Cowley, 1981; Cowley & Lockwood, 1992). For IMF $B_y > 0$, the flux is added more toward postmidnight in the northern lobe and more toward premidnight in the southern lobe. This leads to an asymmetric pressure distribution and sets up the convection pattern described above. Different explanations of how the observed asymmetries arise in the closed magnetosphere have been suggested. It was first proposed that the asymmetry is introduced when open field lines with asymmetric footpoint reconnect in the tail (e.g., Cowley, 1981). Later, an alternative scenario was put forward by Khurana et al. (1996) and expanded upon by Tenfjord et al. (2015). They argued that the asymmetric pressure distribution and associated plasma flows will affect closed field lines directly. A key difference between the two scenarios is the response time of the magnetospheric system to changes in the solar wind forcing, as a more rapid response is expected from the latter mechanism. Several recent studies, using both modeling and observations, have shown that the magnetosphere responds within tens of minutes to polarity changes of the IMF $B_y$ component (Case et al., 2018; Tenfjord et al., 2015, 2017, 2018). This is consistent with the time scales expected by the latter mechanism, which strongly suggest that lobe pressure plays a major role in inducing asymmetries in the closed magnetosphere. For a more comprehensive description of how an IMF $B_y$ component induce asymmetries in the magnetospheric system, we refer the reader to, for example, Tenfjord et al. (2015).

Other mechanisms can contribute to a $B_y$ component, and hence a north-south asymmetry in the closed magnetosphere: (1) warping of the neutral sheet due to temporal variations in the dipole tilt angle (Fairfield, 1980; Russell & Brody, 1967)—this sets up a positive $B_y$ component in the premidnight sector and a negative $B_y$ component in the postmidnight sector for positive tilt angles, and vice versa for negative tilt angles (Petrukovich, 2009; Tsyganenko et al., 1998); (2) a smaller $B_y$ component at all $Y_{GSM}$ positions, positively correlated with the dipole tilt angle (Petrukovich, 2011); (3) rotation of the magnetotail axis driven by the IMF $B_y$, modulating the effect of the $B_y$ induced by lobe pressure (Cowley, 1981; Fairfield, 1979).

Rather than introducing asymmetries, as suggested by Cowley (1981), observations suggest that the closed flux region in the nightside ionosphere becomes more north-south symmetric for enhanced reconnection in the near-Earth tail. Grocott et al. (2010, 2017) performed superposed epoch studies of the ionospheric convection pattern in both hemispheres during substorms and reported a more north-south symmetric pattern in the nightside auroral zones during the expansion phase, regardless of the prevailing IMF conditions. Reistad et al. (2018) used convection data from the Northern Hemisphere divided into subsets based on both AL index and season and found that the return flow became more symmetric with velocities at dusk and dawn becoming more similar for nearly all subsets as |AL| increased. The return to a more symmetric system is also supported by conjugate auroral imaging. In an event study, Østgaard, Humberset, et al. (2011) showed that the longitudinal displacement of conjugate auroral features were reduced during the expansion phase in two subsequent substorms. In a more recent study, Østgaard et al. (2018) found that the asymmetry of conjugate auroral features reached a minimum after substorms occurring during a geomagnetic storm. Ohma et al. (2018) presented a multicase study of conjugate images from substorms, showing a reduced displacement in several events and an apparent relation between the reduction of asymmetry and increase in tail reconnection rate. Statistical support of reduced asymmetry was also found by Milan et al. (2010), who investigated the auroral evolution during substorms in a superposed epoch analysis. There is therefore ample evidence that the closed magnetosphere becomes more symmetric when tail reconnection is enhanced. The plasma convection inside the polar cap, however, still shows the interhemispheric asymmetry expected from IMF $B_y$ control (Grocott et al., 2010; Reistad et al., 2018). Regardless, Reistad et al. (2018) proposed that the observed reduction of asymmetry in the ionospheric convection could imply reduced asymmetry also in the lobe convection in the magnetotail itself. Several studies (e.g., Caan et al., 1975, 1978; Fairfield & Ness, 1970; Yamaguchi et al., 2004) have shown that the near-Earth lobe pressure increases over several hours leading up to substorms and is reduced in less than 1 hr after substorm onset. If the lobe pressure is responsible for the initial asymmetry, it is reasonable to expect that a reduction of the pressure would lead to a reduction in the asymmetry. Figure 1, taken from Reistad et al. (2018), illustrates this conceptually. For quiet conditions, there is no significant tail reconnection, and the convection is highly asymmetric due to the asymmetric loading when IMF $B_y > 0$. For active conditions, there is significant tail reconnection, which makes the convection more symmetric between the hemispheres, even if the IMF $B_y$ is unchanged.

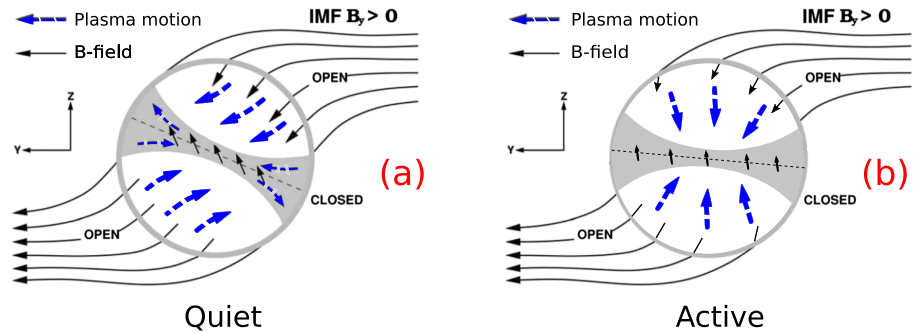

**Figure 1.** Cross section of the magnetosphere seen from the tail when IMF $B_y > 0$, illustrating the suggested interpretation that the magnetotail is more symmetric for enhanced tail reconnection. The figure is taken from Reistad et al. (2018), based on Figure 3a in Liou and Newell (2010). Magnetic field lines are shown as solid black arrows and the plasma convection by the dashed blue arrows. The gray region indicates the closed magnetosphere, and the white regions indicate the lobes. (a) No tail reconnecting, asymmetric flow at both open and closed field lines, and a strong $B_y$ component in the closed magnetosphere. (b) Significant tail reconnection, more symmetric flow at both open and closed field lines, and a weak $B_y$ component in the closed magnetosphere.

If the observed change to a more symmetric state in the ionosphere is indeed a consequence of the entire system returning to a more symmetric state, this should affect the convection pattern also in the magnetotail. The motivation for this study was therefore to investigate whether the change to a more symmetric convection pattern for strong tail reconnection could be observed on open field lines in the magnetospheric lobes. For this purpose, we have used convection measurements from the Electron Drift Instrument (EDI) on board Cluster.

The paper is organized as follows: The EDI data set and the auxiliary data used in this study are presented in section 2. In section 3, we describe the method used to determine the stability of the IMF conditions and how we have mapped the EDI data. We present the mapped convection measurements in section 4, using two different approaches to separate between weak and strong tail reconnection. The results and their impact are discussed in section 5, and we summarize the paper in section 6.

## 2. Data

The Cluster mission (Escoubet et al., 2001) consists of four identical spacecraft flying in a tetrahedron-like formation. During most of the time interval discussed in this paper, Cluster was in a polar elliptic orbit with perigee around 4 $R_E$, apogee around 19.6 $R_E$, and an orbital period of 57 hr. Initially, the apsidal axis of the orbit was near the ecliptic, but due to orbital precession, apogee moved gradually into the Southern Hemisphere. The plane of the orbit is fixed with respect to inertial space, which allows for a complete scan of the magnetosphere every year. Depending on season, Cluster therefore traverses the magnetotail lobes for a substantial part of its orbit.

The convection measurements used in this study are from the EDI on board Cluster. This instrument operates by injecting one or two electron beams into the ambient magnetic field and then detecting the beams after one or more gyrations. The emitted electron beams will only return to the detectors when fired in directions uniquely determined by the plasma drift velocity, and the full velocity vector can then be computed either from the direction of the beams or from the difference in their times of flight. More details about the EDI instrument can be found in Paschmann et al. (1997, 2001) and Quinn et al. (2001).

Cluster has several other instruments that can measure plasma convection, either directly using plasma moments from the Cluster Ion Spectrometry (CIS) experiment (Rème et al., 1997) or the Plasma Electron And Current Experiment (PEACE; Johnstone et al., 1997) or indirectly using electric field measurements from the double probe Electric Field and Wave (EFW) experiment (Gustafsson et al., 1997). Convection data from the EDI instrument have been preferred in this study for a number of reasons. Several of these reasons have already been outlined by Haaland et al. (2007, 2008, 2009, 2017) but are repeated here for convenience. First, the plasma in the lobes and polar cap regions have low density. This can be problematic for both CIS and PEACE, as they need reasonably high count rates to derive accurate distribution functions from which moments are calculated. EDI is not affected by this; in fact, low plasma density is beneficial, as it means less interference when detecting the electron beams. Second, an electrostatic wake can arise around the spacecraft due to spacecraft charging (e.g., Eriksson et al., 2006). Both CIS and PEACE, as well as EFW, suffer from wake effects. EFW measurements will be dominated by the spurious wake field and its influence rather than the convection. For CIS, only ions with energies above the spacecraft charging can reach the instrument. Cold plasma (lower energy part of distribution and dominant population in the lobes) is therefore often missing, and moments will be inaccurate. EDI, on the other hand, is immune to wake effects. Third, solar illumination on the spacecraft can lead to inaccurate measurements in PEACE and EFW. PEACE will be contaminated by low-energy photoelectrons due to this illumination. The derived electric field from EFW will not be accurate if one probe is in sunlight and the other is in darkness, as the photoemission escape of electrons will be asymmetric, with one probe charged more than the other due to photoelectrons. Additionally, EFW is a 2-D experiment, providing the electric field in the spin plane only. The third axis must be determined by assuming that the electric field is perpendicular to the magnetic field, which is not possible if **B** itself is in or near the spin plane. This is in contrast to EDI, which intrinsically provides all three components of the convection velocity.

There are also disadvantages associated with the EDI instrument. It does not provide continuous data, as it fails if the ambient magnetic field is too variable or if there is a high background flux of ∼1-keV electron. Being an active experiment, it can interfere with other experiments on Cluster and charge the spacecraft. EDI is therefore not operated continuously but scheduled so that the interference is minimized. Due to these limitations, measurements from EDI are unavailable in some plasma regimes and will also have data gaps. EDI is, however, ideally suited to determine the plasma velocity with high accuracy in regions with low plasma densities and fairly stable strong field, like in the magnetospheric lobes.

Data from EDI are available from February 2001 for Spacecraft 1, 2, and 3. EDI on Cluster 2 failed in April 2004; the instruments on Spacecraft 1 and 3 are still operational. In this study, we have included data until January 2017. We have used the spin resolution EDI data set (4-s resolution), resampled to 1-min resolution by taking the mean. Convection is a slow process, so time resolution is not crucial.

We have used the solar wind data from the OMNI 1-min data set (King & Papitashvili, 2005) in this study. Data gaps in the data set shorter than 10 min have been linearly interpolated. Periods with longer gaps in the solar wind data have been considered too uncertain and have not been processed further. The AL index is available with 1-min resolution in the OMNI data set and is provided by the World Data Center for Geomagnetism, Kyoto.

## 3. Methodology

### 3.1. Determination of IMF Stability

We have used bias filtering to quantify the value and the stability of the IMF clock angle $\theta_{CA}$ (Haaland et al., 2007). To determine the stability of an IMF measurement, a 30-min interval, starting 20 min prior to the measurement and ending 10 min after, is considered. This time interval is meant to take into account possible errors in the time shift to the bow shock nose as well as some time to set up magnetospheric convection following IMF changes. Each IMF vector within the 30-min interval is normalized, and the mean of all the normalized vectors gives the bias vector. The direction of the bias vector determine $\theta_{CA}$, and the length of the bias vector is a measure of the stability of this angle. If all the vectors point in exactly the same direction, the length of this bias vector becomes unity. If the vectors point in completely random directions, the length will be near zero. We have flagged all measurements where the length of the bias vector is ≥0.93 as stable.

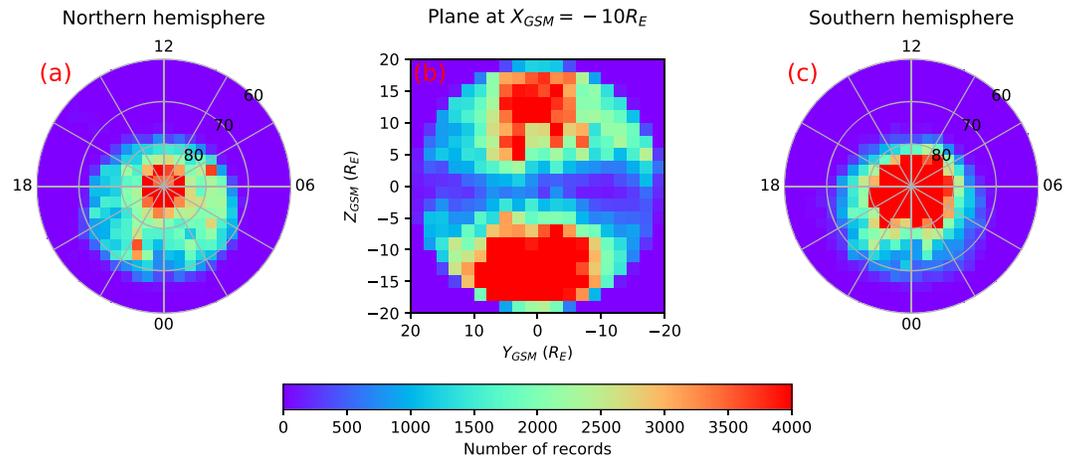

**Figure 2.** Data coverage of the mapped EDI vectors. (a) Number of measurements mapped to the Northern Hemisphere. (b) Coverage of vectors mapped to the plane at $X_{GSM} = -10R_E$. (c) Number of measurements mapped to the Southern Hemisphere.

This limit is slightly lower that the limit used by Haaland et al. (2007) and Haaland et al. (2008). To put the number in context, a linear change in the clock angle of ±36° over a 30-min interval corresponds to a bias vector with a magnitude of 0.931. Using this threshold, about 60% of the data is flagged as stable. Only EDI data from periods of stable IMF have been considered hereafter.

## 3.2. Mapping the EDI Measurements to a Common Plane

The convection of magnetized plasma at one location in space can be mapped along the magnetic field to any location, assuming steady state conditions and equipotential field lines (e.g., Hesse et al., 1997; Maynard et al., 1995). This means that the convection measured by EDI at different altitudes can be mapped to a common plane and that all measurements that map to the plane can be utilized. The velocities $\mathbf{V}_c$ measured by EDI have been mapped to the $YZ_{GSM}$ plane located at $X_{GSM} = -10R_E$. This is the same plane as used in the studies by Noda et al. (2003) and Haaland et al. (2008). The mapped velocity vectors are termed $\mathbf{V}_m$, and hereafter, we use the subscripts $c$ to indicate the position of Cluster and $m$ to indicate the position mapped to $X_{GSM} = -10R_E$. The method used to map the data is identical to the method used by Haaland et al. (2008). For convenience, we summarize the method here: First, the location of Cluster, $\mathbf{x}_{c,0}$, is mapped along the magnetic field to $X_{GSM} = -10R_E$ applying the Tsyganenko 2001 model (Tsyganenko, 2002a, 2002b). In addition to $\mathbf{x}_{c,0}$, we have mapped the point $\mathbf{x}_{c,1}$, which is displaced from $\mathbf{x}_{c,0}$ by a distance $\mathbf{d}_c$ in the direction of $\mathbf{V}_c$. The scaling have been chosen so that the corresponding displacement at $X_{GSM} = -10R_E$, $\mathbf{d}_m = \mathbf{x}_{m,1} - \mathbf{x}_{m,0}$, is $1R_E$. The length of the displacement at the position of Cluster is then given by $d_c = \sqrt{B_m/B_c} \times 1R_E$, where $B_c$ and $B_m$ are the magnitude of the model magnetic field at the two locations. The choice of $d_m = 1R_E$ was chosen by Haaland et al. (2008) as a compromise between uncertainties and distortion in the mapping. The vector $\mathbf{d}_m$ gives the direction of the mapped velocity, and the magnitude is given by $V_m = V_c \times d_m/d_c$. Mapped vectors with magnitude above 50 km/s are removed from the data set, as they can safely be considered as outliers (Haaland et al., 2008).

The spatial data coverage of the mapped data set is displayed in Figure 2. Figure 2a shows the measurements obtained in the Northern Hemisphere mapped to the northern ionosphere, Figure 2b displays coverage of data mapped to the plane at $X_{GSM} = -10R_E$, and Figure 2c displays measurements obtained in the Southern Hemisphere mapped to the southern ionosphere. Figure 2b reveals two distinct regions with high data coverage, one in each lobe. It is also evident that there are more data from the Southern Hemisphere, a consequence of Cluster's orbit precession.

Figure 3 shows the data distribution of the mapped data set. Figure 3a displays the yearly contribution from the different spacecraft and shows that the majority of measurements are obtained prior to 2010. The trend of decreasing number of measured vectors is due to the orbit evolution of the Cluster spacecraft and instrument degradation. Figure 3b shows the seasonal distribution in the two hemispheres. More data are obtained in the Northern Hemisphere autumn months compared with other seasons, which is when Cluster has its apogee in the tail, but there are also contributions from perigee passes in all seasons. Figure 3c displays the altitude distribution of the mapped data. The records from lower altitudes are from perigee passes from all seasons, whereas the records from higher altitudes are from Northern Hemisphere autumn months, when Cluster's apogee is in the tail.

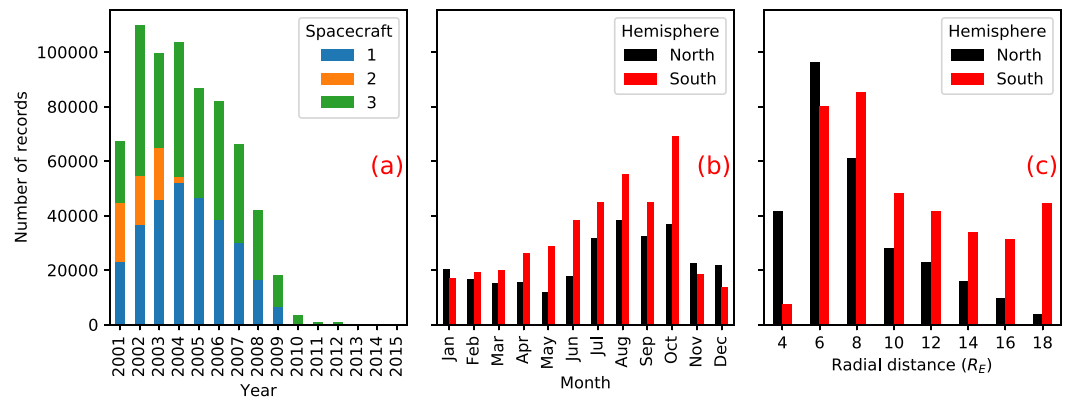

**Figure 3.** Distribution of the mapped EDI vectors. (a) Yearly contribution from the different spacecraft. (b) Seasonal contribution from the Northern (black) and Southern (red) Hemisphere. (c) Contribution from different altitudes for the Northern (black) and Southern (red) Hemisphere.

### 3.3. Isolating the Effect of Near-Earth Tail Reconnection

When the IMF $B_y$ component is positive, the tension force exerted on a newly opened field line drags it toward dawn in the Northern Hemisphere and toward dusk in the Southern Hemisphere. The resulting buildup of pressure initiates convection toward dusk and dawn in the northern and southern lobes, respectively. This is reflected in the ionospheric convection across the polar cap, where the flow have a dawnward component on the dayside and a duskward component on the nightside in the Northern Hemisphere, and vice verse in the Southern Hemisphere (e.g., Haaland et al., 2007), leading to the characteristic banana/orange shaped convection cells. The goal of the present study is to examine whether the part of the convection pattern initiated by the lobe pressure is altered by tail reconnection. To ensure that the correct part of the convection pattern is measured, we have therefore limited the data set to only include measurements that map to the nightside of the dawn-dusk meridian in the ionosphere, but we have not employed any constraint to separate between open an closed field lines. Since EDI usually does not return valid $E$-field data when there is a high background flux of ~1-keV electrons, the convection data obtained by EDI are predominantly from open field lines. This is discussed further in section 5.1. Furthermore, we want to avoid very weak or very extreme solar wind forcing. We have therefore limited the data to only include periods where the solar wind electric field $E_{SW}$ is between 1 and 4 mV/m, where $E_{SW}$ is the product of the radial component of the solar wind velocity and the transverse magnetic field in the IMF, $B_T = \sqrt{B_y^2 + B_z^2}$. We will address the impact of the solar wind forcing further in section 4.3, but note that the exact values used here do not affect the observed trends significantly. In order to investigate how tail activity influences the convection velocities and directions, we use two different selection criteria to divide the data into subsets:

- Approach 1 is to use the value of the AL index to sort the data into activity levels. The results are presented in section 4.1.
- Approach 2 is to separate the data based on substorm phases. The results are presented in section 4.2.

## 4. Observations and Processing

### 4.1. Lobe Convection for Different Levels of the AL Index

Auroral activity (as reflected in the AL index) is strongly correlated with reconnection associated fast flows in the near-Earth magnetotail (Angelopoulos et al., 1994; Baumjohann et al., 1990) and can therefore be used as a proxy of near-Earth tail reconnection. We have therefore sorted the data into two subsets determined by the value of the AL index at the time of each measurement: quiet conditions when AL > −50 and active when AL < −50. This is the same strategy as applied by Reistad et al. (2018) when studying the convection in the ionosphere, but they used three subsets, dividing data with AL < −50 nT into two subsets. The value of −50 nT used here is near the median value of the AL index in the time period considered in this study (−43 nT). We have further divided the data into four clock angle bins; northward and $B_y$ dominated ($\theta_{CA} \in [45°, 90°]$ and $\theta_{CA} \in [−90°, −45°]$) and southward and $B_y$ dominated ($\theta_{CA} \in [90°, 135°]$ and $\theta_{CA} \in [−135°, −90°]$). The convection pattern at $X_{GSM} = −10R_E$ when IMF $B_z > 0$ is displayed in Figure 4 and when IMF $B_z < 0$ in Figure 5. The mapped vectors have been binned into $2 \times 2R_E$ bins, and the vectors displayed in the two figures are the plain mean within each bin. Only bins containing two or more measurements are included in the statistics. The color of each vector indicates the number of measurements in each bin. The left column in both figures is for IMF $B_y > 0$, and the right column is for IMF $B_y < 0$. The upper row is for quiet conditions and the lower row for active conditions.

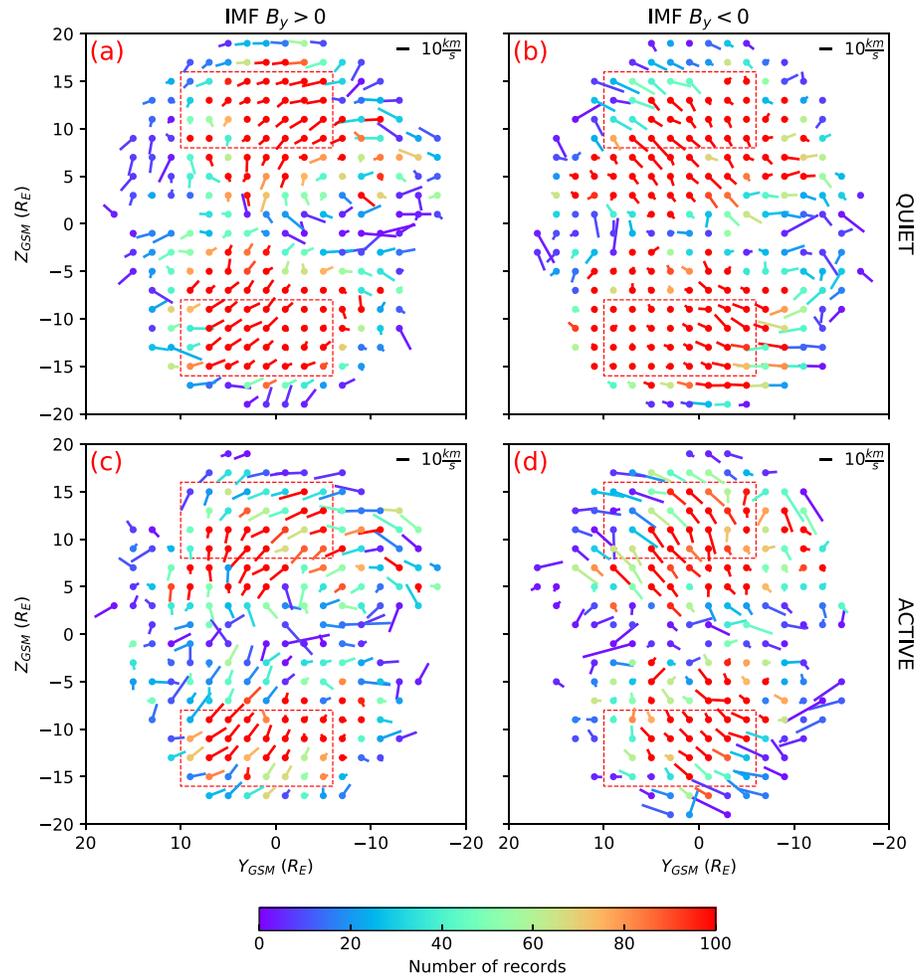

**Figure 4.** Convection pattern when IMF $B_z > 0$, seen from the tail. All vectors are mapped to the plane $X_{GSM} = -10R_E$. Vector scale of 10 km/s is given in the upper-right corner of each panel and color indicate number of records in each bin. Data within the red boxes are used in the statistics in subsequent figures. (a) Quiet (AL > −50) and IMF $B_y > 0$. (b) Quiet (AL > −50) and IMF $B_y < 0$. (c) Active (AL < −50) and IMF $B_y > 0$. (d) Active (AL < −50) and IMF $B_y < 0$.

In order to investigate the direction and magnitude of the large-scale lobe convection, we focus on two regions, one in the northern lobe and one in the southern lobe, spanning $−6R_E < Y_{GSM} < 10R_E$ and $8R_E < |Z_{GSM}| < 16R_E$. These two regions are the same size as the two regions considered by Haaland et al. (2008) but shifted $2R_E$ toward dusk. The rationale behind this shift is because the magnetotail is not symmetric about midnight ($Z_{GSM} = 0$ axis) but rather shifted into the premidnight sector. The average location of substorm onset in the ionosphere is at about 23 MLT (Frey et al., 2004; Liou, 2010), but there is considerable variability in this location. The occurrence frequency of bursty bulk flows associated with tail reconnection is also higher in the premidnight sector (e.g., Angelopoulos et al., 1994; Raj et al., 2002). This means that the average location of the reconnection site is in the premidnight sector. Aberration caused by Earth's orbital motion can contribute to this shift. Simulation efforts by Lu et al. (2016) suggest that the Hall effect is stronger at the duskside as a result of higher ion temperature, thinner current sheet, and smaller normal magnetic field, which also contribute to the shift. It is also possible that nonuniform ionospheric conductance contributes to this dawn-dusk asymmetry (Lotko et al., 2014). The two regions are indicated by the red boxes in Figures 4 and 5.

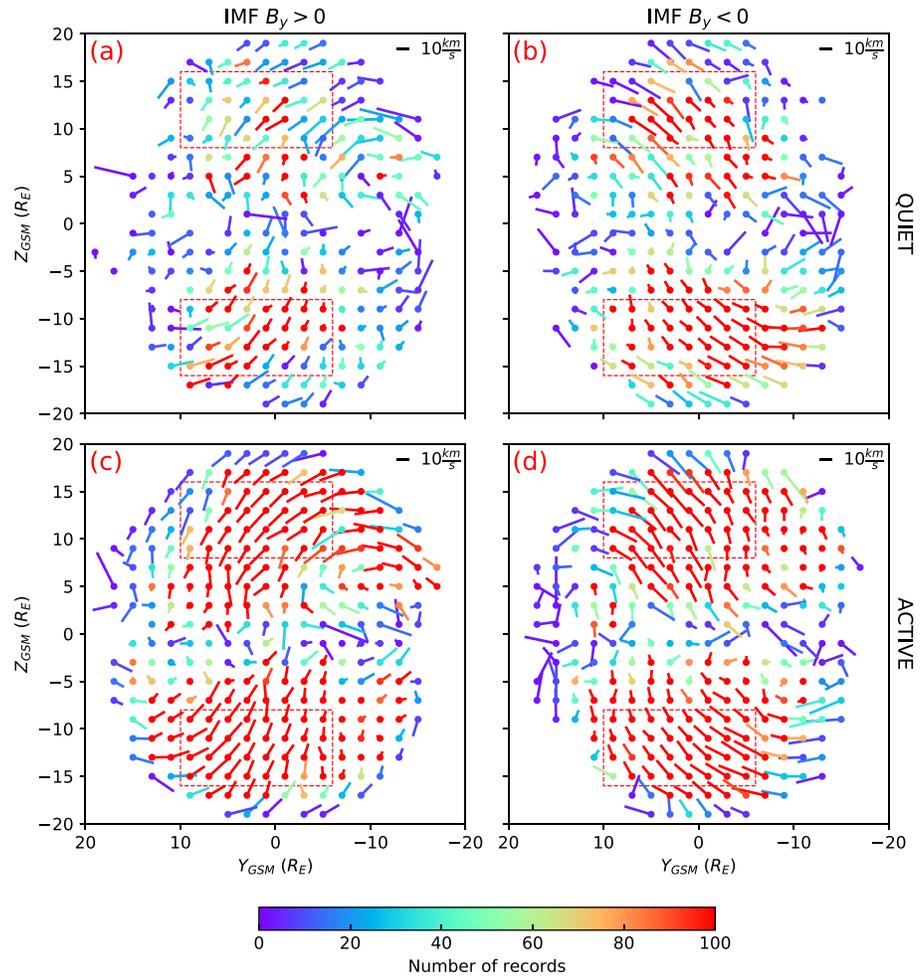

**Figure 5.** Same as Figure 4 but for IMF $B_z < 0$.

For each subset of the data, we estimate the average convection vector $\mathbf{V}$ inside the two boxes. From Figures 4 and 5, it is clear that the data are unevenly distributed within the boxes. To compensate for this, we have done a weighted average, using the number of records in the $2 \times 2R_E$ bins to determine the weight of each vector. There are 32 bins within each box. Each vector in bin $i$ gets the weight

$$w_i = \frac{\bar{n}}{n_i}, \tag{1}$$

where $n_i$ is the number of records in that bin and $\bar{n}$ is the average number of records in all the bins within the box containing data. Using this formula, each bin is weighted equally, which means that this is equivalent to taking the mean of the mean vectors inside the box.

In order to estimate the error of $\mathbf{V}$, we calculate the covariance matrix of the vector components, using the weights given in equation (1) in the calculation. The two eigenvalues of the covariance matrix, $\sigma_{max}^2$ and $\sigma_{min}^2$, describes the maximum and minimum variance of the data, respectively, and their associated eigenvectors gives the directions of the spread. Both eigenvalues are very similar in all subsets in this study, so we have assumed the largest variance in all directions. The standard error of the mean perpendicular to the direction of $\mathbf{V}$ is then given as

$$\sigma = \sqrt{\frac{\sigma_{max}^2}{n}}, \tag{2}$$

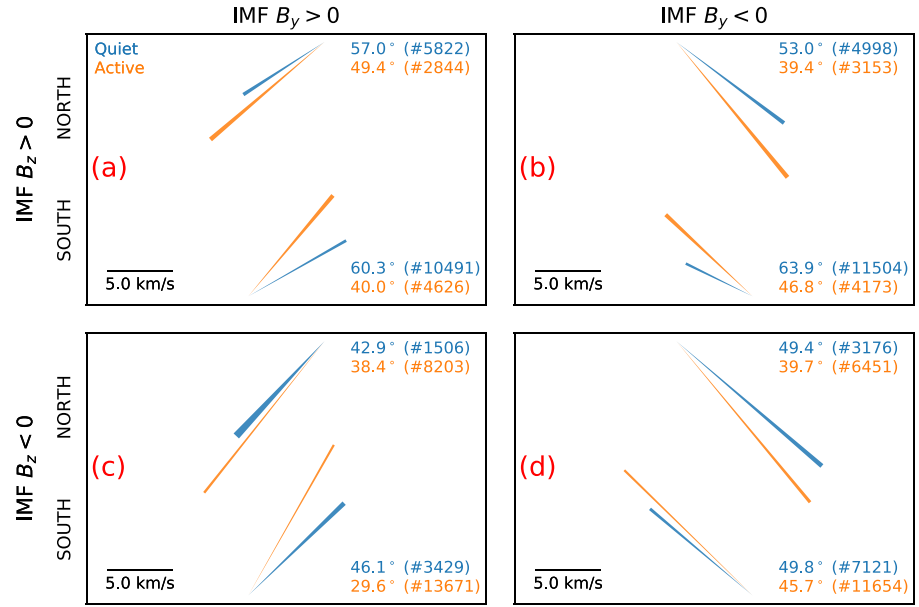

**Figure 6.** Average convection in the central lobes for quiet (blue) and active (orange) conditions, seen from the tail. Upper (lower) vectors in each panel are from the northern (southern) lobe. The numbers in the panels indicate the angle between each vector and the $Z_{GSM}$ axis, and the width of each vector indicate the standard error. The number in parenthesis is the number of data points in each average. We note that the flow becomes more north-south aligned during active periods; that is, asymmetry is reduced. (a) IMF $B_y > 0$ and IMF $B_z > 0$. (b) IMF $B_y < 0$ and IMF $B_z > 0$. (c) IMF $B_y > 0$ and IMF $B_z < 0$. (d) IMF $B_y < 0$ and IMF $B_z < 0$.

where $n$ is the total number of records inside the box. Using small angle approximation, the angular error of the direction of $\mathbf{V}$ is then

$$\sigma_{\mathbf{V}} = \frac{\sigma}{|\mathbf{V}|}. \tag{3}$$

The average convection in the two central lobes is displayed in Figure 6. Figures 6a and 6c are when IMF $B_y > 0$, and Figures 6b and 6d are when IMF $B_y < 0$. As in Figures 4 and 5, the convection is seen from the tail. Blue and orange vectors indicate quiet and active conditions, respectively. The vectors point in the direction of the mean flow and the lengths are proportional to the magnitude of the convection. The numbers by the vectors indicate the angle between the the velocity vector and the $Z_{GSM}$ axis, where 0° is directly toward the plasma sheet. The total number of records in the statistics are indicated by the number in the parenthesis, and the width of the vectors indicate the uncertainty estimated using equation (3). From the figure, it is clear that the magnitude of the convection increases in the active subsets. Furthermore, we observe that the magnitude of the convection is larger for IMF $B_z < 0$ compared to IMF $B_z > 0$. The quiet time subsets have the largest angle, which means that the convection becomes more north-south aligned, that is, the asymmetry reduces, during active conditions.

Another approach to determine the direction of the convection is to disregard the magnitude of the mapped vectors and only consider the direction of normalized vectors. Each mapped vector points in an angle $\theta_i$ from the $Z_{GSM}$ axis, which means that the average of the normalized vectors is given as

$$\bar{\mathbf{R}} = \left[ \langle \sin \theta_i \rangle, \langle \cos \theta_i \rangle \right], \tag{4}$$

where the angle brackets indicate the weighted average, using the weights given by equation (1). This vector is equivalent to the bias vector described above. The average angle is given by

$$\bar{\theta} = \tan^{-1} \frac{\langle \sin \theta_i \rangle}{\langle \cos \theta_i \rangle}, \tag{5}$$

and the length of $\bar{\mathbf{R}}$ is given by

$$\bar{R} = \sqrt{\langle \sin \theta_i \rangle^2 + \langle \cos \theta_i \rangle^2}. \tag{6}$$

This last quantity, $\bar{R}$, is equivalent to the magnitude of the bias vector and is thus a measure of the circular spread. In fact, Fisher (1993) defines the normalized circular variance as

$$\sigma_{circ}^2 = 1 - \bar{R}. \tag{7}$$

Further, $\bar{R}$ is the first central trigonometric moment relative to $\bar{\theta}$ and can be expressed as

$$\bar{R} = m_1 = \langle \cos\left(\theta_i - \bar{\theta}\right)\rangle. \tag{8}$$

Similarly, the second central trigonometric moment is defined as

$$m_2 = \langle \cos 2\left(\theta_i - \bar{\theta}\right)\rangle. \tag{9}$$

From the two trigonometric moments, another measure of circular spread can be defined as (Fisher, 1993)

$$\hat{\delta} = \frac{1 - m_2}{2m_1^2}. \tag{10}$$

This quantity is referred to as the sample circular dispersion and is an important quantity when comparing $\bar{\theta}$ from different samples. The circular standard error $\sigma_{\bar{\theta}}$ is given as

$$\sigma_{\bar{\theta}} = \sqrt{\frac{\hat{\delta}}{n}} \tag{11}$$

and is hereby used to estimate the error of $\bar{\theta}$.

Other methods to estimate the spread of vector quantities exist. Förster et al. (2007) define the normalized variance of the magnitude as

$$\sigma_{mag}^2 = 1 - \frac{\langle |\mathbf{V}|\rangle^2}{\langle |\mathbf{V}|^2\rangle} \tag{12}$$

an the total variance of the vector as

$$\sigma_{tot}^2 = 1 - \frac{|\langle \mathbf{V}\rangle|^2}{\langle |\mathbf{V}|^2\rangle}. \tag{13}$$

Using these on the EDI data in the central lobes, we find that $\sigma_{mag}$ is small and that the total variance and circular variance are very similar in all subsets.

The angles between the convection and the $Z_{GSM}$ axis inferred from both methods described above are plotted in Figure 7. The black and red lines are from the northern and southern lobes, respectively. The bright lines are the angles of the average velocity vectors, and the dim lines the angles calculated by equation (5). From the figure, it is clear that the values and errors estimated by both methods are quite similar and that both methods show the same trend. The figure highlights that the plasma flow becomes more oriented toward the plasma sheet for increased AL index, which means that the convection is more north-south aligned. We also note that the value of the angles are similar in the Northern and Southern Hemisphere.

The average solar wind forcing and dipole tilt angles for the different subsets in Figures 6 and 7 are given in Table 1. The table shows that both the average solar wind electric field $\langle E_{SW}\rangle$ and the average IMF clock angle $|\langle\theta_{CA}\rangle|$ are higher in the active subsets compared to the quiet time subsets. Higher values of $E_{SW}$ and $|\theta_{CA}|$ are associated with an increase in the dayside reconnection rate (e.g., Newell et al., 2007), which means that more flux is loaded during active time subsets. This is to be expected, as unloading of flux by tail reconnection must, on average, be balanced by loading of flux from dayside reconnection, but it also means that the external forcing on the magnetosphere is different for the quiet and active time subsets. The increase in $\langle E_{SW}\rangle$ means that more flux is added asymmetrically, which should *increase* the asymmetry in the active subsets and therefore reduce the signatures observed in Figures 6 and 7. The increase in $|\langle\theta_{CA}\rangle|$ when IMF $B_z < 0$, on the other hand, means that the flux is added, on average, closer to the $Z_{GSM}$ axis. This could contribute to a reduction in angle of the convection. We will show in section 4.3 that this is not the main mechanism responsible for the observed changes. Table 1 also displays the average dipole tilt angle $\langle\psi\rangle$, just above zero, in the different subsets. There are no apparent bias for IMF $B_y > 0$, but some difference between the quiet and active time subsets for IMF $B_y < 0$.

In summary: Using AL as proxy, we see that the convection becomes more north-south aligned during periods of strong reconnection in the near-Earth tail.

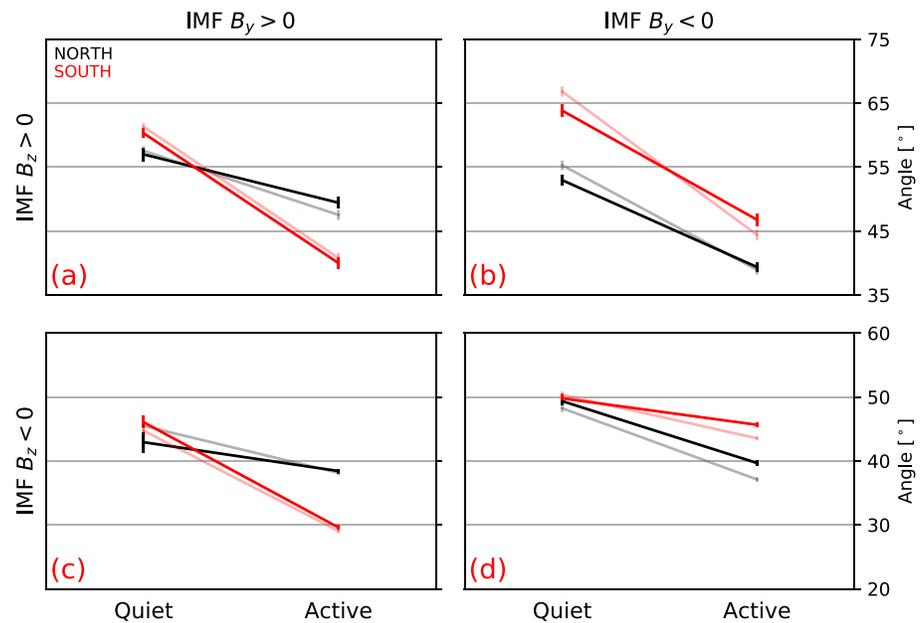

**Figure 7.** Angle between the convection and the $Z_{GSM}$ axis for quiet and active conditions. Black and red indicate northern and southern lobe, respectively. The bright lines are inferred from the average convection vectors and the dim lines from equation (5). (a) IMF $B_y > 0$ and IMF $B_z > 0$. (b) IMF $B_y < 0$ and IMF $B_z > 0$. (c) IMF $B_y > 0$ and IMF $B_z < 0$. (d) IMF $B_y < 0$ and IMF $B_z < 0$.

## 4.2. Lobe Convection for Different Substorm Phases

The closure of lobe flux, as well as the rate of the strongly related bursty bulk flows, peaks in the substorm expansion phase (e.g., Baumjohann et al., 1991; Milan et al., 2007). Juusola et al. (2011) have shown that the bursty bulk flows stay at an elevated level also well into the recovery phase. This means that we can use substorm phases to group the convection data, as the expansion and recovery phases are associated with increased reconnection in the near-Earth tail. Forsyth et al. (2015) have developed a technique for determining substorm phases automatically, termed Substorm Onsets and Phases from Indices of the Electrojet (SOPHIE). The study includes lists where the technique has been applied to the SML index (Gjerloev, 2012). In short, the SOPHIE technique works by applying a low pass filter to the electrojet index and then finding its time derivative. Expansion phase is identified when dSML/dt is below a specific threshold and recovery phase is identified when dSML/dt is above a specific threshold. Other intervals are identified as "possible growth" phase. The data set also undergoes postprocessing, where for instance short periods of growth phase identified between expansion and recovery are included in one of the substorm phases and the identified phases are adjusted to match the unfiltered SML index. The list provided by Forsyth et al. (2015) with the least stringent criteria to identify the expansion and recovery phases has been used.

The convection in the central lobes sorted by the phases identified by SOPHIE is presented in Figure 8. The figure has the same format as Figure 6, and the vectors and their uncertainties have been calculated in a similar manner. The number of substorms indicated in the figure is the total number of unique expansion phases that contribute to the statistics. The average magnitude of the convection is larger for the expansion and recovery phases compared to the possible growth phase. The direction of the convection is more directed toward the plasma sheet for seven vectors in both the expansion phase and the recovery phase subsets, with a single exception seen in Figure 8d, where the direction is unchanged in the Southern Hemisphere. The figure shows no consistent distinction between the expansion and recovery subsets, neither in directions nor magnitudes. Figure 9 displays the angles between the convection and the $Z_{GSM}$ axis, similarly to Figure 7 in the previous section. The figure highlights that the angles are reduced in nearly all subsets associated with enhanced near-Earth tail reconnection. We note that the angle is reduced in all subsets when only the directions of the vectors are considered (dim lines). The average solar wind forcing and dipole tilt angle for the different subsets are given in Table 2. The table shows that the differences between the subsets are in the same sense as discussed in the previous section, but smaller.

**Table 1**
*Average Solar Wind Forcing for the Different Clock Angle Bins*

| | | IMF $B_y > 0$ | | | IMF $B_y < 0$ | | |
|---|---|---|---|---|---|---|---|
| | | $\langle E_{SW} \rangle$ (mV/m) | $\langle \theta_{CA} \rangle$ (°) | $\langle \psi \rangle$ (°) | $\langle E_{SW} \rangle$ (mV/m) | $\langle \theta_{CA} \rangle$ (°) | $\langle \psi \rangle$ (°) |
| IMF $B_z > 0$ | Quiet | 2.0 | 66 | 4 | 1.9 | −68 | 1 |
| | Active | 2.4 | 72 | 3 | 2.3 | −74 | 7 |
| IMF $B_z < 0$ | Quiet | 1.8 | 107 | 4 | 1.7 | −108 | −1 |
| | Active | 2.3 | 114 | 2 | 2.3 | −113 | 6 |

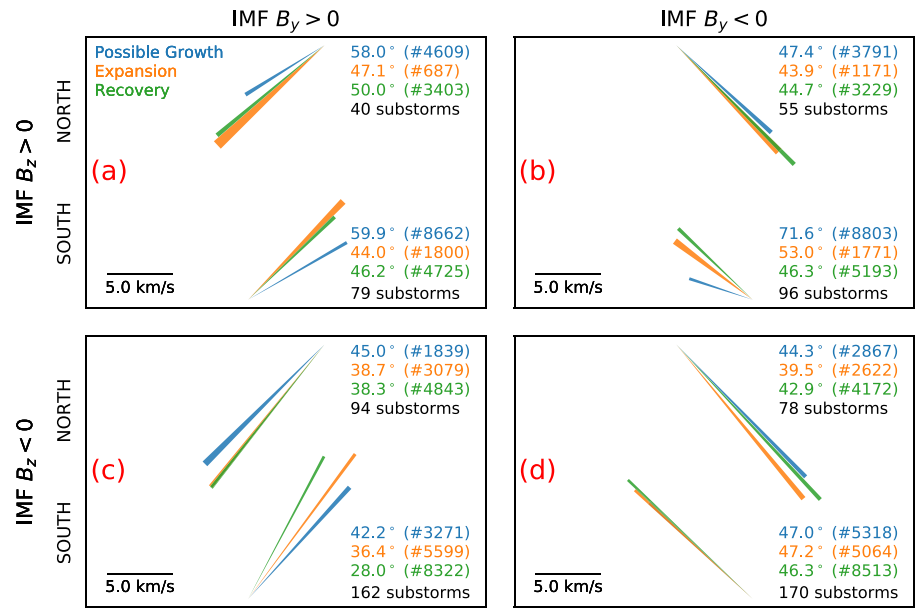

**Figure 8.** Average convection in the central lobes for possible growth (blue), expansion (orange), and recovery (green) phase based on SOPHIE. Upper (lower) vectors in each panel are from the northern (southern) lobe. The numbers in the panels indicate the angle between each vector and the $Z_{GSM}$ axis, and the width of each vector indicates the standard error. The number in parenthesis is the number of data points in each average, and the number of substorms is the total number of unique expansion phases that contribute to the averages. (a) IMF $B_y > 0$ and IMF $B_z > 0$. (b) IMF $B_y < 0$ and IMF $B_z > 0$. (c) IMF $B_y > 0$ and IMF $B_z < 0$. (d) IMF $B_y < 0$ and IMF $B_z < 0$.

Using identified substorms, we once again see a shift toward a more north-south aligned convection in the subsets associated with strong reconnection in the near-Earth tail.

### 4.3. Role of Solar Wind Forcing

As discussed in sections 4.1 and 4.2, the external forcing is different in the quiet and active subsets, which is shown in Tables 1 and 2. In order to verify that the changes observed in section 4.1 and 4.2 are not caused by changes in the external solar wind forcing, we sort the data according to the external driving. We assume that the Northern Hemisphere is a flipped mirror of the Southern Hemisphere and improve statistics by combining data from both hemispheres and both polarities of the IMF $B_y$. This has been done by flipping the measurements in the Northern Hemisphere across the $Z_{GSM} = 0$ axis and by flipping the measurements in the Southern Hemisphere when IMF $B_y > 0$ and in the Northern Hemisphere when IMF $B_y < 0$ across the $Y_{GSM} = 2R_E$ axis (center of the boxes in Figures 4 and 5). This combination of measurements assumes IMF $B_y$ control on the plasma flow. While this is not necessarily the case on closed field lines during substorms (e.g., Grocott et al., 2017), the convection on open field lines inside the polar cap shows the expected IMF $B_y$ asymmetry also during active and substorm conditions (Grocott et al., 2010, 2017; Reistad et al., 2018). The persistence of the IMF $B_y$ control is also evident in Figures 6 and 8. The data are then divided into bins according to $E_{SW}$ at the time of each measurement, using a bin size of 1 mV/m. In each bin, we separate the data based on the substorm phases determined by SOPHIE, as this method provides most equal external forcing inside each $E_{SW}$ bin.

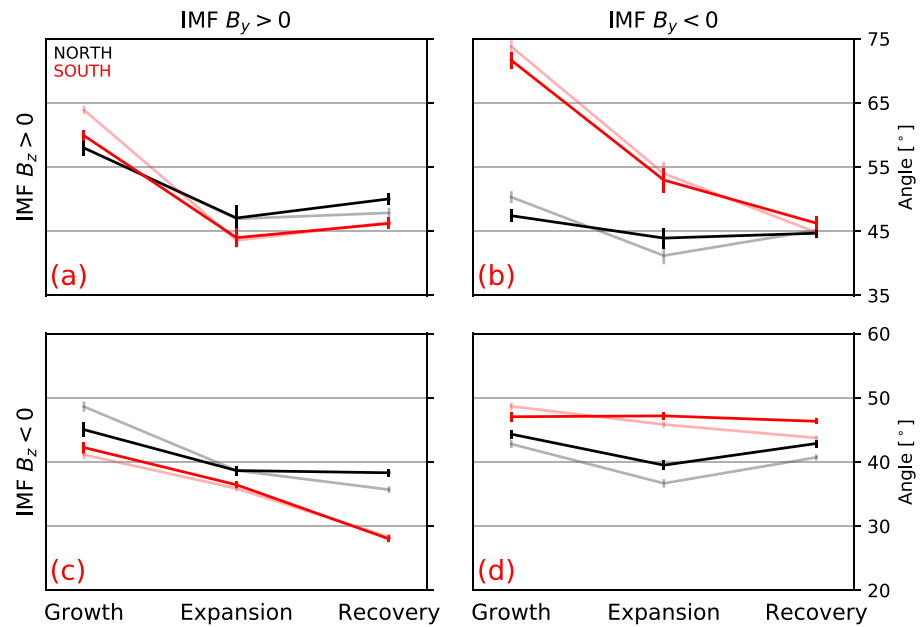

**Figure 9.** Angle between the convection and the $Z_{GSM}$ axis for the different substorm phases. Black and red indicate northern and southern lobe, respectively. The bright lines are inferred from the average convection vectors and the dim lines from equation (5). (a) IMF $B_y > 0$ and IMF $B_z > 0$. (b) IMF $B_y < 0$ and IMF $B_z > 0$. (c) IMF $B_y > 0$ and IMF $B_z < 0$. (d) IMF $B_y < 0$ and IMF $B_z < 0$.

The result of this combined statistics for $|\theta_{CA}| \in [90°, 135°]$ is shown in Figure 10, where the average $V_y$ component is displayed in Figure 10a, the average $V_z$ is displayed in Figure 10b, and the angle between the average convection direction and the $Z_{GSM}$ axis is shown in Figure 10c. Figure 10c displays that the angle for expansion and recovery phase is consistently lower than the angle in the growth phase. The reduced angle is associated with an increase in the $V_z$ component (flow toward the plasma sheet) with no corresponding increase in the $V_y$ component. Figure 10a also shows that the magnitude of the $V_y$ component increases rapidly as $E_{SW}$ increases and is affected to a much larger degree than the $V_z$ component. The direction of the convection is therefore sensitive to the level of solar wind forcing, where the angle increases as $E_{SW}$ increases. Since the average solar wind forcing is higher in the expansion and recovery phase subsets compared to the growth phase subsets, this effect will reduce the signature seen in sections 4.1 and 4.2 and could explain the small change observed in Figure 8d. The trends seen in Figures 10b and 10c are the same for $|\theta_{CA}| \in [45°, 90°]$.

To further examine the clock angle dependence, we have divided the bin with $E_{sw}$ between 1 and 2 mV/m into clock angle bins of 15°. This is by far the bin with most records, and the only bin where the data are evenly divided between quiet and active conditions. The expansion and recovery phase subsets have been combined to improve data coverage, as the two subsets behave very similarly (Figures 8 and 10). The results are displayed in Figure 11, which has the same format as Figure 10. In Figure 11a, we see that the $V_y$ component is about equal in both subsets, whereas Figure 11b shows that the $V_z$ component is consistently higher for the combined expansion and recovery phase subset. This means that the lobe convection is consistently more north-south aligned during substorm conditions (Figure 11c), even when we consider a limited interval of solar wind forcing and narrow clock angle bins.

**Table 2**
*Average Solar Wind Forcing for the Different Clock Angle Bins*

| | | IMF $B_y > 0$ | | | IMF $B_y < 0$ | | |
|---|---|---|---|---|---|---|---|
| | | $\langle E_{SW} \rangle$ (mV/m) | $\langle \theta_{CA} \rangle$ (°) | $\langle \psi \rangle$ (°) | $\langle E_{SW} \rangle$ (mV/m) | $\langle \theta_{CA} \rangle$ (°) | $\langle \psi \rangle$ (°) |
| | Possible growth | 2.0 | 66 | 5 | 1.9 | −67 | 2 |
| IMF $B_z > 0$ | Expansion | 2.3 | 71 | 1 | 2.1 | −74 | 6 |
| | Recovery | 2.3 | 71 | 2 | 2.1 | −71 | 4 |
| | Possible growth | 1.9 | 108 | 4 | 1.7 | −109 | 1 |
| IMF $B_z < 0$ | Expansion | 2.3 | 115 | 2 | 2.3 | −114 | 5 |
| | Recovery | 2.3 | 112 | 2 | 2.2 | −111 | 4 |

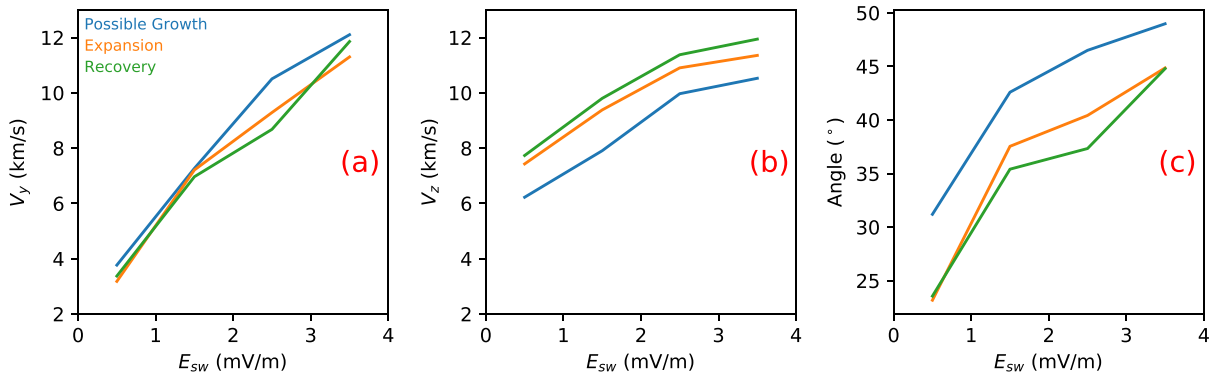

**Figure 10.** Average convection in the central lobes for different levels of solar wind forcing using the combined data set and $|\theta_{CA}| \in [90°, 135°]$. The blue, orange, and green curves display possible growth phase, expansion phase, and recovery phase, respectively. (a) Average $V_y$ component. (b) Average $V_z$ component. (c) Angle between the average convection and the $Z_{GSM}$ axis.

Figures 10 and 11 thus confirm that the observed changes in the average convection in the lobes are not caused by changes in the external solar wind forcing.

## 5. Discussion

The above results show that the convection in the lobes becomes more north-south aligned in the subsets inferred to be associated with strong near-Earth tail reconnection. The change of the average convection is due to an increase of the average convection toward the plasma sheet, without a corresponding increase in the transverse ($V_y$) component. Even though the direction is more north-south aligned compared to the quiet subsets, the convection is still showing the asymmetry expected from IMF $B_y$ control. Before we move on to discuss the implications of the presented results, we summarize some of the uncertainties and assumptions in the data and methodology and how they can affect our results.

### 5.1. Robustness of the Results

- Measurement errors: In the lobes, with its stable and fairly strong magnetic field, the EDI principle ensures very accurate full 3-D convection measurements. EDI is not affected by spacecraft wake effects due to spacecraft charging or photoemissions that can affect measurements from double probe instruments and particle detectors. We infer that measurement errors have negligible effect on the results.

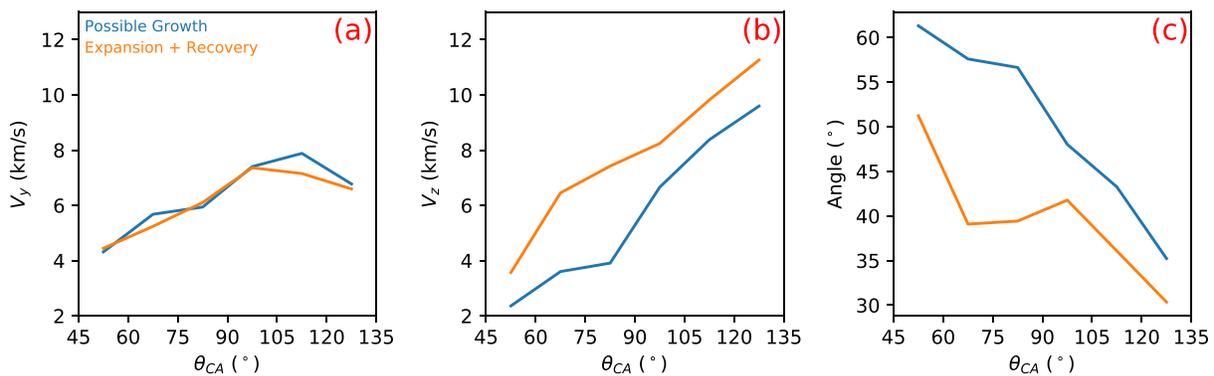

**Figure 11.** Average convection in the central lobes for different clock angles using the combined data set and $E_{SW}$ between 1 and 2 mV/m. The blue line displays "possible growth" phase and the orange line displays expansion and recovery phase combined. (a) Average $V_y$ component. (b) Average $V_z$ component. (c) Angle between the average convection and the $Z_{GSM}$ axis.

- Location of the obtained convection measurements: We have not employed any criteria to determine whether a measurement is obtained in the lobes or in the plasma sheet, as EDI rarely returns valid $E$-field data in the closed flux regions. To check the validity of this approach, we have considered the plasma beta for each data record, which is defined as the plasma pressure divided by the magnetic pressure. The plasma pressure used are from the CIS experiment, whereas the magnetic pressure is inferred from the EDI measurements. Data from CIS are available for about 82% of the EDI measurements. For 99% of the EDI data points within our defined reference boxes, $\beta < 0.03$, which clearly shows that the measurements were obtained in the lobes (Baumjohann et al., 1988, 1989).

- Mapping of convection measurements: The mapping from position of Cluster to the $X_{GSM} = -10R_E$ plane assumes no parallel electric fields. This is a reasonable assumption in the collisionless plasma environment of the lobes. There is also uncertainty associated with the accuracy of the magnetic field model, but the work by Woodfield et al. (2007) suggests that the Tsyganenko model is accurate in a statistical sense. It is also assumed that the convection is in the $YZ_{GSM}$ plane at $X_{GSM} = -10R_E$, hence that the magnetic field only has a $B_x$ component. Values from the Tsyganenko model at the mapped locations indicate that this is a good approximation; $B_x$ is on average a factor of 5 larger than the other components. The results from Noda et al. (2003) also indicate that the convection is predominantly in the $YZ$-plane here.

- Statistical spread in measurements: Our results are based on a large number of EDI measurements over a long time period, from which we derive averages. The measurements obviously posses a large statistical variation. We tried several methods to characterize the "average," several methods to estimate the statistical errors in these averages and different sizes and positions of area we use to calculate averages over. The results are consistent and all show the same trend—a more north-south aligned convection for strong near-Earth tail reconnection.

- Defining periods with tail reconnection: Since we investigate if magnetotail reconnection is the process that reduces asymmetries, we need to identify periods with tail reconnection. Admittedly, we do not have direct in situ measurements of tail reconnection, but the relation between our proxies, the AL index and substorm phases, and reconnection in the near-Earth tail, is well established (Angelopoulos et al., 1994; Baumjohann et al., 1990, 1991). Although this relation is not one to one, the proxies should be able to identify intervals with either high or low likelihood of near-Earth tail reconnection. The AL index is largely due to internal magnetospheric dynamics (Newell et al., 2007), but a part is directly related to solar wind forcing. It can therefore be assumed that the AL index have two sources, one part driven directly by the solar wind and one part driven by the unloading of magnetic energy stored in the magnetotail (Blanchard & McPherron, 1995; Bargatze et al., 1985; McPherron et al., 1988). As seen in Figure 10c, increased solar wind forcing leads to more east-west aligned convection. The observed change to a more north-south aligned convection for high |AL| must therefore be associated with unloading of energy in the magnetotail, which is related to enhanced near-Earth tail reconnection. It is also relevant to point out that the proxies used in this study is not expected to reflect the reconnection rate in the distant tail, which means that we compare intervals with weak or strong reconnection in the near-Earth tail. We note that there is an overlap between the quiet and growth subsets, and likewise between the active and expansion/recovery subsets, from the two methods.

- The role of IMF: We observe a more north-south aligned convection during periods inferred to be associated with strong tail reconnection. It is reasonable to ask whether this reduction can be attributed to differences in the external solar wind/IMF forcing. However, our analysis clearly shows that IMF cannot explain the changes and that the symmetry restoration must be caused by tail reconnection.

## 5.2. Interpretation of the Results

As described in the introduction, observations of the ionospheric convection suggest that the convection pattern in the auroral zones becomes more north-south symmetric during periods with strong reconnection in the near-Earth tail (Grocott et al., 2010, 2017; Reistad et al., 2018). While the convection pattern can become almost completely north-south symmetric in the closed flux region, the observed convection inside the polar cap still shows the asymmetry expected from IMF $B_y$ control. This is consistent with the observations in this study, where we observe a change to a more north-south aligned convection in the lobes for strong reconnection in the near-Earth tail, but still showing the asymmetry expected from IMF $B_y$ control.

Reduced asymmetry is also frequently observed in conjugate auroral images during substorm expansion phase (Ohma et al., 2018; Østgaard et al., 2004, Østgaard, Humberset, et al., 2011, 2018). Ohma et al. (2018) showed that this was associated with an enhancement of the tail reconnection rate. A more symmetric system is also consistent with the observations of Owen et al. (1995), who studied the orientation of the distant plasma sheet boundary layer. When the IMF is dominated by a $B_y$ component, they found that the tail is significantly more twisted for northward IMF compared to southward IMF. Further, they found that the distribution of tail twists becomes narrower for higher levels of geomagnetic activity compared to lower levels. This means that the tail is able to twist more for lower levels of activity, but becomes more symmetric for enhanced activity.

Substorm reconnection commences at closed plasma sheet field lines, but as the substorm evolves, the near-Earth reconnection progresses into open field lines. Open lobe flux is therefore closed (e.g., Milan et al., 2007), which could affect the convection in the lobes. As the newly closed flux is transported away from the reconnection region, the pressure reduces in the central magnetotail and lobe plasma converges toward the reconnection region. This means that the convection toward the plasma sheet increases, as observed, and could account for the more north-south aligned convection. The transport of magnetic flux toward the plasma sheet also depletes the pressure in the two lobes. Caan et al. (1975, 1978) and Yamaguchi et al. (2004) have shown that the lobe pressure increases in the hours before substorm onset and then rapidly decreases within 1 hr. The buildup of pressure prior to onset is associated with the loading of magnetic flux from dayside reconnection without an immediately corresponding increase in the tail reconnection rate. The magnetosphere will expand as flux is added and subject it to a larger pressure component of the flowing magnetosheath plasma and cause the observed increase in pressure (Fairfield & Ness, 1970). During the substorm expansion phase, open magnetic flux is efficiently removed from the lobes (e.g., Caan et al., 1975). The decrease in pressure following substorm onset can be interpreted as a result of increase in the tail reconnection rate, which dissipates magnetic energy stored in the tail and deflates the magnetosphere. Milan et al. (2008) have shown observational evidence that such a direct link between changes in open magnetic flux in the polar cap and the lobe field strength exists. If asymmetries in the magnetosphere are produced by asymmetric lobe pressure, it is possible that a reduction of lobe pressure could contribute to less asymmetric convection.

In our analysis, we do not consider the location of substorm onset, and we have therefore averaged substorms with onsets in different MLT sectors. However, there is considerable variability in onset location (Frey et al., 2004), and Grocott et al. (2017) have shown that the morphology of the convection in the auroral zones depends on onset location. The convection can differ substantially for atypically early or atypically late onsets occurring during the same IMF conditions; the convection throat moves toward dusk in both hemispheres for early onsets, and the convection throat moves toward dawn in both hemispheres for late onsets. If our statistics is a mix of early and late onsets, we could also be averaging a mix of dawnward and duskward flows associated with the onset location, leading to a more north-south aligned flow on average, but with higher variability. We do, however, find that the circular variance (and thus the circular standard deviation; Fisher, 1993) consistently decrease for increasing activity. This means that there is less variation in the direction of the convection in the subsets associated with enhanced near-Earth tail reconnection, indicating that onset location does not influence the result in the central lobes. It is, however, possible that such effects could influence the convection closer to the plasma sheet. It is also worth noting that atypically early or late onset are far less common than onsets at typical MLT locations (e.g., Grocott et al., 2010), which means that the latter dominates the statistics.

## 6. Conclusions

In this paper we have used convection data from EDI mapped to a common plane in the lobes to study the convection for different levels of tail activity when the IMF has a dominating east-west ($B_y$) component. The findings can be summarized as the following:

1. The average convection in the central lobes becomes more north-south aligned for high values of |AL| compared to low values of |AL|.
2. The average convection is more north-south aligned for nearly all vectors in the expansion and recovery phase subsets compared to the "possible growth" phase subsets.
3. By combining data from both hemispheres and both polarities of the IMF $B_y$, we find that the observed differences in the average convection persisted even for narrow intervals of solar wind forcing and IMF clock angles. This suggests that the changes cannot be attributed to differences in the external forcing.

4. These results indicate that strong reconnection in the near-Earth tail significantly affects the average plasma convection also in the magnetotail lobes, which becomes more north-south aligned for enhanced reconnection.

5. The more symmetric pattern associated with enhanced tail reconnection implies that the asymmetric pressure balance in the magnetotail lobes has been altered.

**Acknowledgments**

This study was supported by the Research Council of Norway under Contract 223252/F50 (CoE). The Cluster EDI and CIS data are provided by ESA's Cluster Science Archive and can be accessed online (at https://csa.esac.esa.int/csa-web/). We acknowledge use of NASA/GSFC's Space Physics Data Facility's OMNIWeb service, and OMNI data, which can be accessed online (at http://omniweb.gsfc.nasa.gov). We thank the World Data Center for Geomagnetism, Kyoto, for providing the AL index, which is included in the OMNI data set.